\documentclass[conference]{IEEEtran}
\usepackage{tabularx}
    \newcolumntype{L}{>{\raggedright\arraybackslash}X}
\usepackage{comment}
\usepackage{amsmath}

\usepackage{amssymb}
\usepackage{gensymb}
\usepackage{graphicx}
\usepackage{epstopdf} 
\usepackage{color}
\usepackage[bookmarks=false,hidelinks]{hyperref}
\usepackage{enumerate}
\usepackage{multirow}
\usepackage{cite}
\usepackage{multicol}
\usepackage{threeparttable}
\usepackage{subfig}
\usepackage{float}
\usepackage{mathtools}
\usepackage{array}
\newcolumntype{P}[1]{>{\centering\arraybackslash}p{#1}}
\newcolumntype{M}[1]{>{\centering\arraybackslash}m{#1}}
\usepackage{nopageno}
\usepackage{pifont}
\newcommand{\cmark}{\ding{51}}
\newcommand{\xmark}{\ding{55}}

\usepackage{url}

\usepackage[numbers]{natbib} 
\usepackage{xcolor}

\def\BibTeX{{\rm B\kern-.05em{\sc i\kern-.025em b}\kern-.08emT\kern-.1667em\lower.7ex\hbox{E}\kern-.125emX}}

\usepackage{graphicx}
\usepackage{nicefrac}
\usepackage{siunitx}
\usepackage{array,framed}
\usepackage{booktabs}
\usepackage{graphicx}
\usepackage{graphics}

\usepackage{
  tikz,
  pgfplots,
  pgfplotstable
}
\usepackage{hyperref}

\usetikzlibrary{
  shapes.geometric,
  arrows,
  external,
  pgfplots.groupplots,
  matrix
}

\pgfplotsset{compat=1.9}

\usepackage{mathtools,}

\DeclareMathAlphabet{\mathcal}{OMS}{cmsy}{m}{n}

\DeclareGraphicsExtensions{%
    .png,.PNG,%
    .pdf,.PDF,%
    .jpg,.mps,.jpeg,.jbig2,.jb2,.JPG,.JPEG,.JBIG2,.JB2}

\usepackage{xparse}
\newcommand{\bnm}{\begin{newmath}}
\newcommand{\enm}{\end{newmath}}

\newcommand{\bea}{\begin{eqnarray*}}%
\newcommand{\eea}{\end{eqnarray*}}%

\newcommand{\bne}{\begin{newequation}}
\newcommand{\ene}{\end{newequation}}

\newcommand{\bal}{\begin{newalign}}
\newcommand{\eal}{\end{newalign}}

\newenvironment{newalign}{\begin{align}%
\setlength{\abovedisplayskip}{4pt}%
\setlength{\belowdisplayskip}{4pt}%
\setlength{\abovedisplayshortskip}{6pt}%
\setlength{\belowdisplayshortskip}{6pt} }{\end{align}}

\newenvironment{newmath}{\begin{displaymath}%
\setlength{\abovedisplayskip}{4pt}%
\setlength{\belowdisplayskip}{4pt}%
\setlength{\abovedisplayshortskip}{6pt}%
\setlength{\belowdisplayshortskip}{6pt} }{\end{displaymath}}

\newenvironment{newequation}{\begin{equation}%
\setlength{\abovedisplayskip}{4pt}%
\setlength{\belowdisplayskip}{4pt}%
\setlength{\abovedisplayshortskip}{6pt}%
\setlength{\belowdisplayshortskip}{6pt} }{\end{equation}}

\newcounter{ctr}

\newcounter{mytable}
\def\mytable{\begin{centering}\refstepcounter{mytable}}
\def\endmytable{\end{centering}}

\newcounter{myfig}
\def\myfig{\begin{centering}\refstepcounter{myfig}}
\def\endmyfig{\end{centering}}

\newlength{\saveparindent}
\newlength{\saveparskip}
\newcommand{\E}{{\rm I\kern-.3em E}}

\renewcommand{\eqref}[1]{\mbox{Equation~(\ref{#1})}}

\def \part {part}

\renewcommand{\paragraph}[1]{\vspace*{6pt}\noindent\textbf{#1}\;}

\def \blackslug{\hbox{\hskip 1pt \vrule width 4pt height 8pt
    depth 1.5pt \hskip 1pt}}
\def \qed{\quad\blackslug\lower 8.5pt\null\par}

\newcounter{mynote}[section]

\newcommand\ignore[1]{}

\newcounter{rcnote}[section]

\newcounter{mrnote}[section]

\newcounter{fknote}[section]

\newcounter{anote}[section]

\DeclareMathSymbol{\mlq}{\mathord}{operators}{``}
\DeclareMathSymbol{\mrq}{\mathord}{operators}{`'}

\newcommand{\rhf}[2]{R_{f, \gamma}}

\DeclareDocumentCommand{\edist}{o o}{
  \ensuremath{
    \IfNoValueTF{#1}{{d}}{{\sf d}(#1,#2)}
  }
}

\newcommand{\olrk}[1]{\ifx\nursymbol#1\else\!\!\mskip4.5mu plus 0.5mu\left(\mskip0.5mu plus0.5mu #1\mskip1.5mu plus0.5mu \right)\fi}

\NewDocumentCommand{\indseq}{ O{1} O{r} }{{#1}\ldots {#2}}

\usepackage{atbegshi}
\AtBeginDocument{\AtBeginShipoutNext{\AtBeginShipoutDiscard}}

\begin{document}

\begin{abstract}
Machine learning (ML)-based methods have recently become attractive for detecting security vulnerability 
exploits. Unfortunately, state-of-the-art ML models like long short-term memories (LSTMs) and 
transformers incur significant computation overheads. This overhead makes it infeasible to deploy them in 
real-time environments. We propose a novel ML-based exploit detection model, ML-FEED, that enables
highly efficient inference without sacrificing performance. 
We develop a novel automated technique to extract 
vulnerability patterns from the Common Weakness Enumeration (CWE) and Common Vulnerabilities and Exposures (CVE)
databases. This feature enables ML-FEED to be aware of the latest cyber weaknesses. 
Second, it is not based on the traditional approach of classifying sequences of application
programming interface (API) calls into exploit categories. Such traditional methods that process entire 
sequences incur huge computational overheads. Instead, ML-FEED operates at a finer granularity and predicts 
the exploits triggered by every API call of the program trace. Then, it uses a state table to update the 
states of these potential exploits and track the progress of potential exploit chains. ML-FEED also employs 
a feature engineering approach that uses natural language processing-based word embeddings, frequency 
vectors, and one-hot encoding to detect semantically-similar instruction calls. Then, it updates the states 
of the predicted exploit categories and triggers an alarm when a vulnerability fingerprint executes. Our 
experiments show that ML-FEED is $72.9\times$ and $75,828.9\times$ faster than 
state-of-the-art lightweight LSTM and transformer models, respectively. We trained and tested ML-FEED on 
79 real-world exploit categories. It predicts categories of exploit in real-time with 98.2\% precision, 
97.4\% recall, and 97.8\% F1 score. These results also outperform the LSTM and transformer baselines.
In addition, we evaluated ML-FEED on the attack traces of CVE vulnerability exploits in three popular Java libraries and detected 
all three reported critical vulnerabilities in them.
\end{abstract}

\title{ML-FEED: Machine Learning Framework for Efficient Exploit Detection}

\author{
\IEEEauthorblockN{Tanujay Saha}
\IEEEauthorblockA{\textit{Electrical and Computer Engineering}\\
\textit{Princeton University}\\
NJ, USA\\
tsaha@princeton.edu}

\and

\IEEEauthorblockN{Tamjid Al Rahat}
\IEEEauthorblockA{\textit{Electrical and Computer Engineering}\\
\textit{University of California Los Angeles}\\
CA, USA\\
tamjid@ucla.edu}

\and

\IEEEauthorblockN{Najwa Aaraj}
\IEEEauthorblockA{\textit{Cryptography Research Center}\\
\textit{Technology Innovation Institute}\\
Abu Dhabi, UAE\\
najwa@tii.ae}

\and

\IEEEauthorblockN{Yuan Tian}
\IEEEauthorblockA{\textit{Electrical and Computer Engineering}\\
\textit{University of California Los Angeles}\\
CA, USA\\
yuant@ucla.edu}

\and

\IEEEauthorblockN{Niraj K. Jha}
\IEEEauthorblockA{\textit{Electrical and Computer Engineering} \\
\textit{Princeton University}\\
NJ, USA \\
jha@princeton.edu}
}


\maketitle




\section{Introduction}
\label{sec:intro}

Digital platforms worldwide have been facing increasingly sophisticated malicious campaigns directed at them. 
Most cyber-attacks are executed by exploiting system vulnerabilities. 
Vulnerabilities are weaknesses in a system that hackers can exploit in multiple ways. Vulnerability scanning involves detecting such potential weaknesses. This requires administrator-level access to the system resources. Exploit detection requires more sophistication than vulnerability detection. This is because system administrators often do not grant permission to exploit detection software to access internal resources.

Researchers have devoted significant efforts towards addressing the challenge of detecting 
vulnerabilities and exploits, developing popular methods like fuzzing, 
symbolic analysis, 
and rule-based testing. 
Although these methods are quite popular, they do have some drawbacks. A significant limitation is that most such
methods require manual definitions of attack signatures and patterns. Manual description of rules limits the 
efficiency of these methods when dealing with large codebases~\cite{aaraj2008dynamic}. 
Moreover, these traditional vulnerability detection techniques suffer from a large number of false positives, performance issues, and inability to identify the vulnerability type~\cite{yamaguchi2014modeling}. Researchers have overcome these drawbacks by integrating machine learning (ML), more specifically deep learning, into vulnerability detection techniques.

Using ML requires minimal manual effort. Thus,
it expedites the process. State-of-the-art ML methods, like long-short term memories (LSTMs) and transformers, classify application programming interface (API) call sequences obtained from program execution traces as benign or malignant. They 
can even predict the exploit type. However, these models incur huge computation overheads. This diminishes their usefulness. 

This article presents ML-FEED (Machine Learning Framework for Efficient Exploit Detection) 
that combines ML and non-ML methods to achieve highly efficient exploit classification of execution traces.
Next, we discuss the drawbacks of state-of-the-art techniques and how ML-FEED overcomes them.
Transformers analyze the entire sequence of instruction calls at every timestep during inference,
resulting in huge computation overheads. LSTMs use representations of the sequence history to reason about 
the entire sequence at every timestep. Although these representations help them capture semantic information 
in the sequence, handling these representations incurs large computation overheads. Therefore, unlike
traditional ML-based sequence models, ML-FEED does not analyze the entire sequence of instruction calls
at once. It first predicts the exploits that an instruction call may trigger. Then, it uses this
prediction to update the states of the predicted exploits in a state table. This state table keeps track
of the current execution state of every exploit in our threat model. When the state of one or more exploits in the table represents a vulnerability fingerprint, it raises the alarm for a potential attack. Our experiments 
show that ML-FEED is faster than the state-of-the-art lightweight LSTM (transformer) model by $72.9\times$
($75,828.9\times$) and $4.5\times$ ($2500.6\times$) more compact.

Besides designing a lightweight model for efficient inference, we also design a general framework for
extracting threat intelligence from vulnerability databases like Common Weakness Enumeration (CWE) and
Common Vulnerabilities and Exposures (CVE). We use this extracted intelligence to construct an exploit sequences database, which we further use
to train our ML model. This methodology enables us to update our ML model with the latest exploits, thus
mitigating the impact of a lack of sufficient data consisting of heterogeneous vulnerabilities.

The novelties of our approach are as follows.
\begin{enumerate}
    
    \item We propose a hybrid system comprising ML classifiers and state tables to classify exploit chains efficiently. 
    
    \item We analyze individual API calls instead of entire API call sequences. This analysis at a 
finer granularity enables ML-FEED to demonstrate much higher efficiency than state-of-the-art sequence models.
    

    \item We propose a methodology to automatically extract intelligence from vulnerability databases and 
create an exploit sequence database to update ML-based exploit detection systems with the latest exploits.
\end{enumerate}

The article is organized as follows. Section \ref{sec:background} provides the background information 
necessary for understanding the rest of the article. 
Section \ref{sec:Methodology} describes the methodology for exploit classification. Section \ref{sec:eval} details the results of our experiments. Section 
\ref{sec:related_work} discusses related work in this area. 
Section \ref{sec:conclusion} concludes the article. 

\section{Background}
\label{sec:background}
This section discusses background material for program representations and ML models.


\subsection{Call Graph}
\begin{figure}[h]
     \centering
     \includegraphics[scale=0.55]{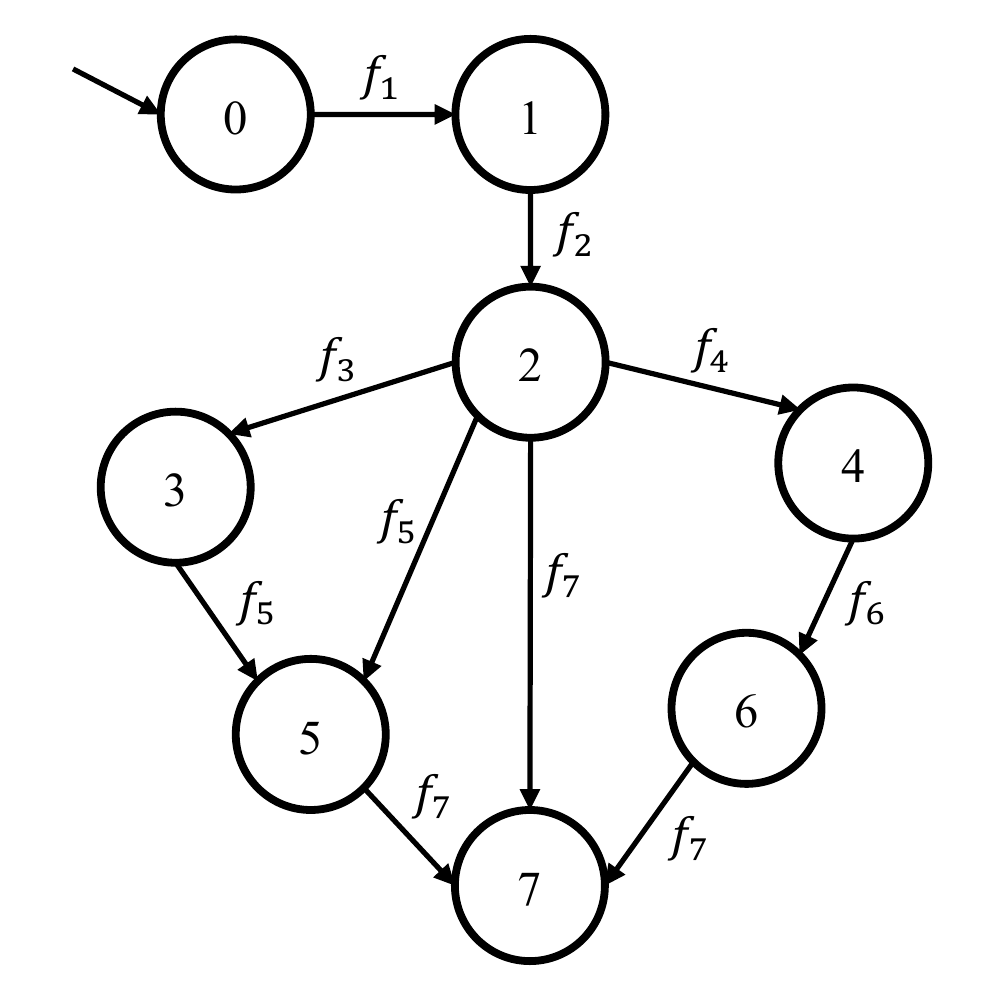}
     \caption{Example of a call graph.}
     \label{fig:callgraph}
\end{figure}

A call graph is a control-flow graph representing the calling relationships between functions in a program. Fig.~\ref{fig:callgraph} shows an example of a call graph where each node represents a function, and each edge represents the calling relationship between the functions. For example, if a function $f_1$ calls $f_2$ during the program flow, the calling relationship is represented by a directed edge ($f_1$,$f_2$). 


\subsection{System Dependence Graph}
We use a \textit{System Dependence Graph} (\texttt{SDG})~\cite{cerberus2022} as the inter-procedural graph representation of programs to extract the vulnerable API sequences for training our model. An SDG $\mathcal{G} = (V, E^{c}, E^{d})$ is a three-tuple, where $V$ represents program statements, $E^{c}$ encodes control dependency edges, and $E^{d}$ denotes data dependency 
edges.  For nodes $X$ and $Y$ in an SDG, $Y$ is control-dependent on $X$ if,
during execution, $X$ can directly affect whether $Y$ is executed.
Also, $X$ is data-dependent on $Y$ if $Y$ is an assignment and  
 $X$ can access the value assigned in $Y$. 
Since these data and control dependencies are computed for individual program procedures (or functions), they do not account for inter-procedural control and data relationships. As a result, they cannot determine the transitive dependencies by default. Therefore, we augment the SDG with information on inter-procedural relationships. SDG edges also represent the direct dependence between the call site and the called procedure, which allows us to determine the transitive dependencies due to a call. 

To construct the SDG, a call graph and pointer analysis is simultaneously performed over the entire program. A call graph is a control-flow graph that represents the calling relationships between functions in a program. Each node represents a function and each edge represents the calling relationship between the functions. Since we target object-oriented programs, we use the Class Hierarchy Analysis (CHA) \cite{dean1995optimization}
call graph algorithm implemented in the Wala framework. On the other hand, pointer analysis 
is a static code analysis technique that determines which pointers (i.e., references) can point to which variables or storage locations. Since we target Java-based program in our analyses, we use an Andersen-style pointer analysis, 
which is available in the Wala toolchain.


\begin{figure}[h]
    \centering
    \vspace{-0.2cm}
    \includegraphics[scale=0.7]{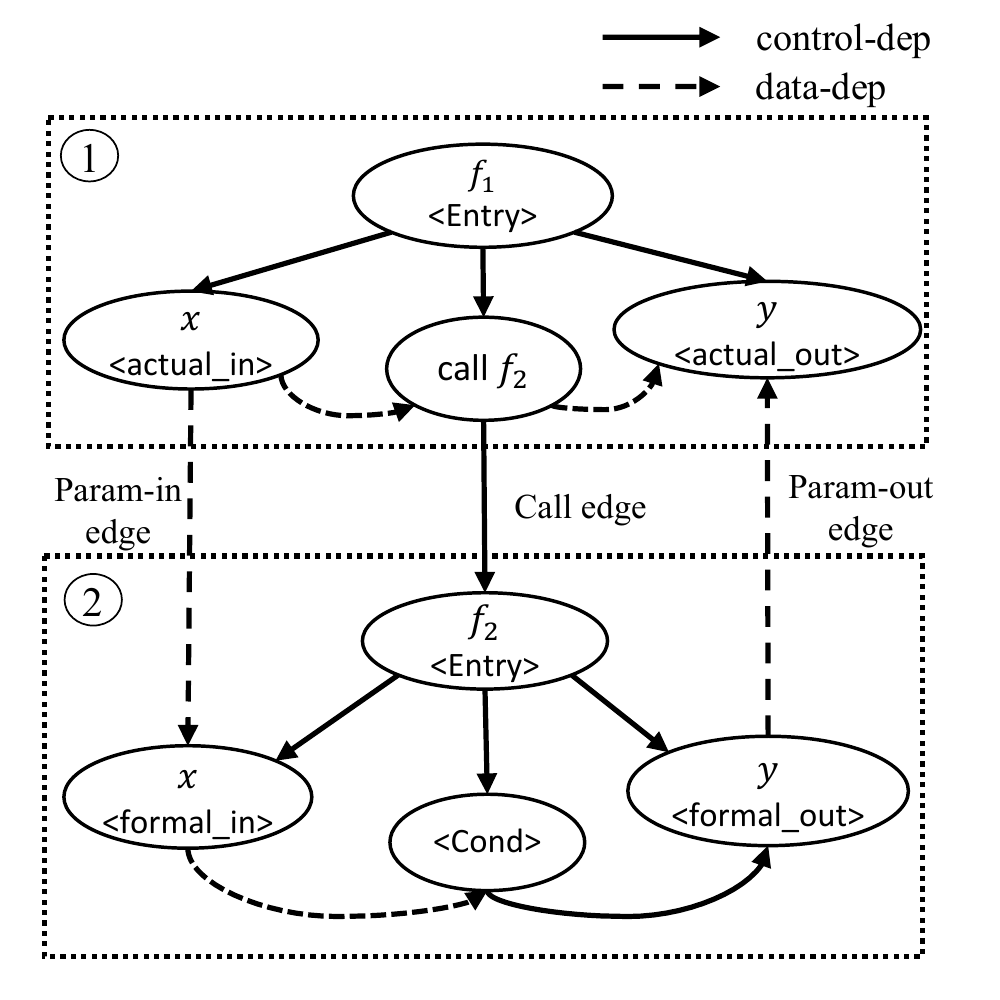}
    \caption{Example of an SDG for call graph nodes 1 and 2. Here, an SDG node represents a 
statement and an edge represents the control or data dependency between program statements.}
    \label{fig:sdg}
\end{figure}

An SDG creates three additional edges to handle function calls: (1) call edge, (2) parameter-in edge, and
(3) parameter-out edge. A call edge connects the node at the call site to the entry node of the called
procedure. Parameter-in edges connect the actual-in parameter nodes to the formal-in parameter nodes of
the called function, and parameter-out edges connect the formal-out nodes to the actual-out nodes.
Fig.~\ref{fig:sdg} shows an SDG for two functions, $f_1$ and $f_2$, where $f_1$ calls $f_2$.
When $f_1$ calls $f_2$ during the program flow, a call edge from the call site in caller $f_1$ to the
entry node of callee $f_2$ represents this calling relationship. Also, the arguments passed by the
caller are represented by the \textit{actual\_in} nodes, and the \textit{formal\_in} nodes represent the
arguments received by the callee. The returned value from the caller is represented
by the \textit{formal\_out} node, whereas the value received by the callee represents the \textit{actual\_out} 
node. As shown in Fig.~\ref{fig:sdg}, the solid and dotted lines represent the control and data dependencies 
between the SDG nodes, respectively.



\subsection{Sequence Models}
\begin{figure}[h]
\centering
\includegraphics[width=1.05\linewidth,scale=1.2, height =0.66\linewidth]{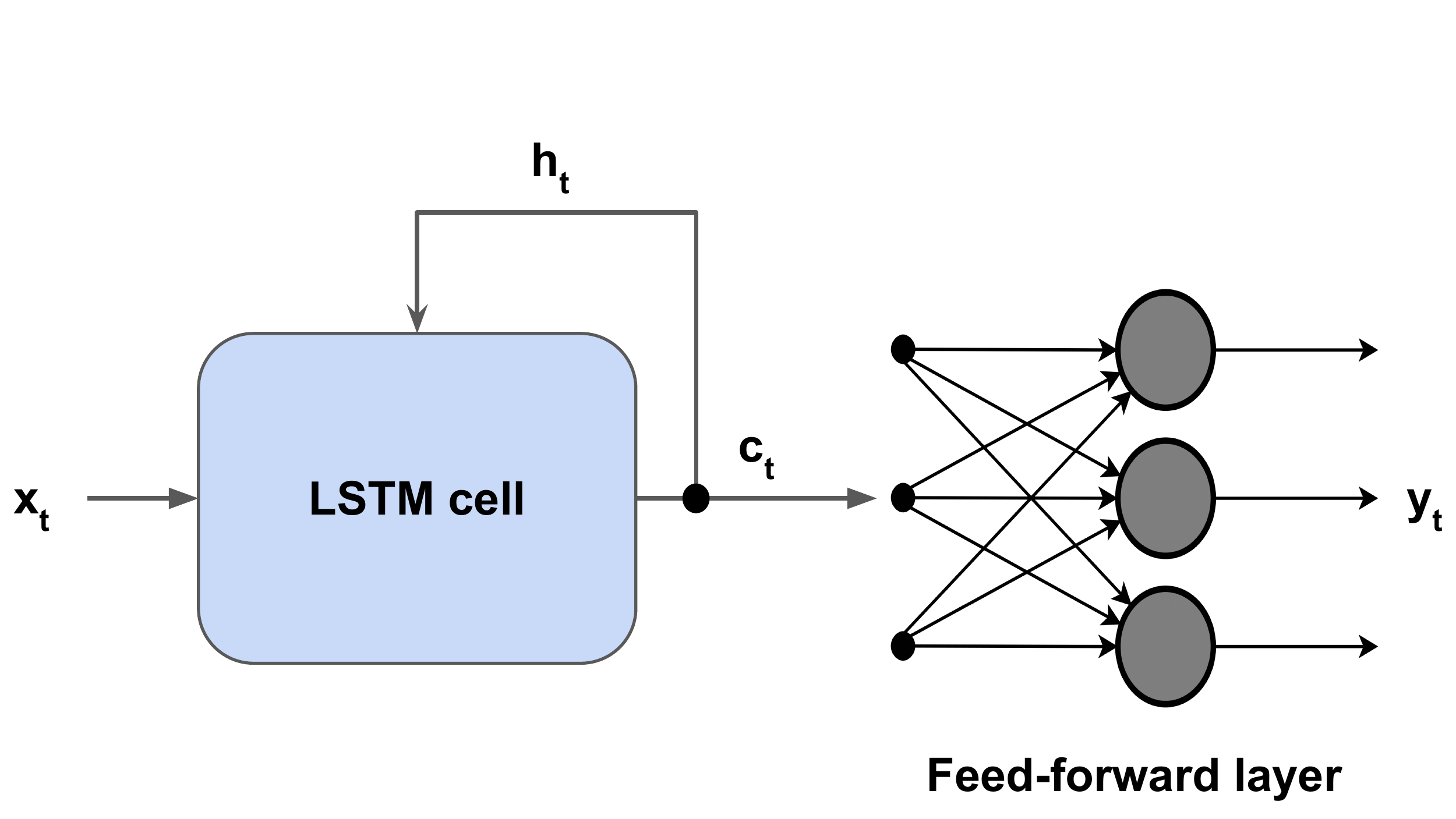}
\caption{The high-level LSTM-based architecture for vulnerability exploit detection. $x_t$, $h_t$, $c_t$, and $y_t$ denote the input feature vector, hidden state, cell state, and multi-label output at time $t$, respectively.} 
\label{fig:lstm-architecture}
\end{figure}

Sequence models are ML-based models that can analyze data with sequential dependencies. Time-series 
data, video and audio clips, and text streams are a few examples of sequential data. A popular deep learning model used for sequential modeling is the LSTM. This is because LSTMs are capable of handling long-range sequential dependencies. Security analysts often use LSTMs and their variants to analyze system-call sequences for vulnerability detection. 
More sophisticated versions of the LSTM, like bidirectional-LSTMs \cite{li2018vuldeepecker} and CNN-LSTMs \cite{8322752}, usually achieve good performance over vanilla LSTM models. 
However, LSTM models require significant computation overheads due to their increased complexities. Therefore, we compare ML-FEED with the state-of-the-art lightweight LSTM model that has reasonable accuracy while taking minimal computation time. As shown in Fig.~\ref{fig:lstm-architecture}, this architecture is a cascade of an LSTM cell followed by a feedforward neural network. 


Transformers are another category of sequence analysis models. 
Transformers have become popular because they are even more successful at capturing long-range dependencies with the help of the attention mechanism. This mechanism 
considers the entire series during inference and determines how to prioritize the different elements of a sequence. Furthermore, their advanced semantic knowledge representation enables sophisticated modeling of semantically-similar functions. 
Although very successful, transformers are often huge models that incur enormous computation overheads. Transformer architectures that have performed well on natural language processing (NLP) tasks have also been successful in system-call sequence analysis. Some of the most popular transformer architectures used for detecting vulnerabilities include the generative pre-trained
(GPT) transformer and its variants
, BERT and BART
. Unlike BERT and BART, the GPT family of transformers is a decoder-based model. It is easier to model our problem statement with a decoder model~\cite{csahin2021malware}. Thus, we compare the performance of ML-FEED with that of GPT-2.
\section{Methodology}
\label{sec:Methodology}

We describe our methodology for efficient exploit detection in this section. 
We begin by extracting vulnerable instruction sequences 
from vulnerability databases and publicly available vulnerability exploit programs. Then, we convert the 
individual instruction calls into labeled feature vectors. These vectors constitute our training data. We 
use our training data to design and train our neural network architecture for the ML-FEED exploit classifier. 

\begin{figure}[h]
   \includegraphics[scale=0.43]{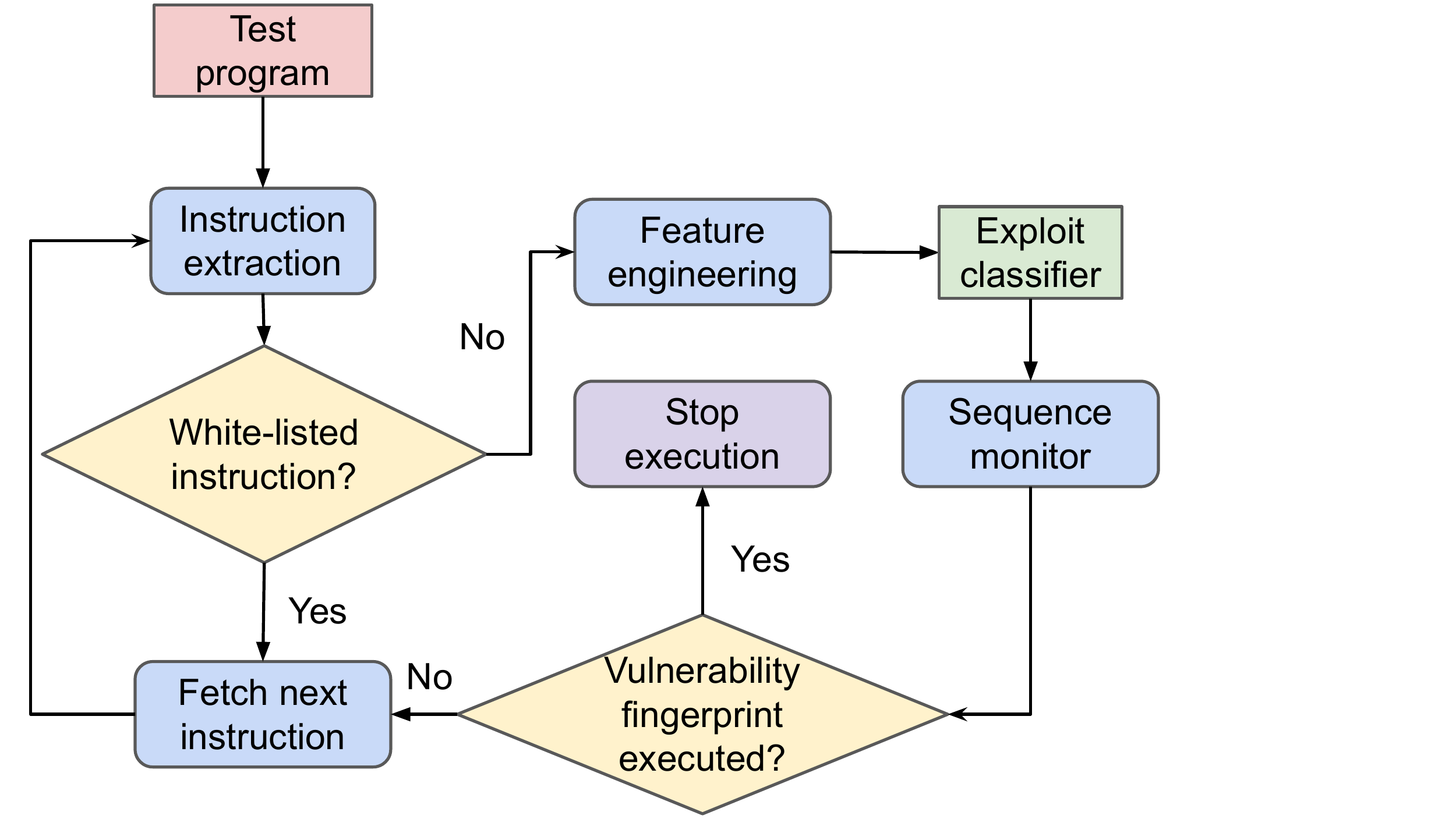}
   \caption{Overview of our inference pipeline}
   \label{fig:overview-inference}
\end{figure}

We give a high-level overview of the ML-FEED inference pipeline in Fig.~\ref{fig:overview-inference}. First, 
we extract instruction calls sequentially from the test program during inference. If the current instruction call is present in the white-list of non-harmful functions, we proceed to analyze the next instruction. Otherwise, we continue to analyze the current function by converting it into a feature vector. The exploit classifier then analyzes this feature vector to predict the potential exploits that the current instruction call can execute. Finally, this list of potential exploits is provided as input to the sequence monitor. The sequence monitor tracks the order of the instruction calls and raises an alarm if the current instruction executes a vulnerability fingerprint. If the current instruction does not implement an exploit, we analyze the next instruction of the test program execution trace.

During inference, ML-FEED raises an alarm when it detects the execution of a CVE/CWE vulnerability fingerprint. Prior ML-based methods (like LSTMs and transformers) and non-ML-based methods (like subgraph matching) for vulnerability detection are not efficient enough to be deployed in real-time. Non-ML-based methods incur a time overhead of $28.57\times$~\cite{aaraj2008dynamic}. As we show later, prior ML-based methods like lightweight LSTMs and transformers incur time overheads of $72.86\times$ and $76178.88\times$, respectively. We compare the efficiency of ML-FEED with prior methods in Section~\ref{sec:sequence-monitor}.

\subsection{Representation of Vulnerable Programs}
\label{sec:Representation_vuln_programs}
\begin{figure*}[h]
    \centering
    \vspace{0.2cm}
    \includegraphics[scale=0.53]{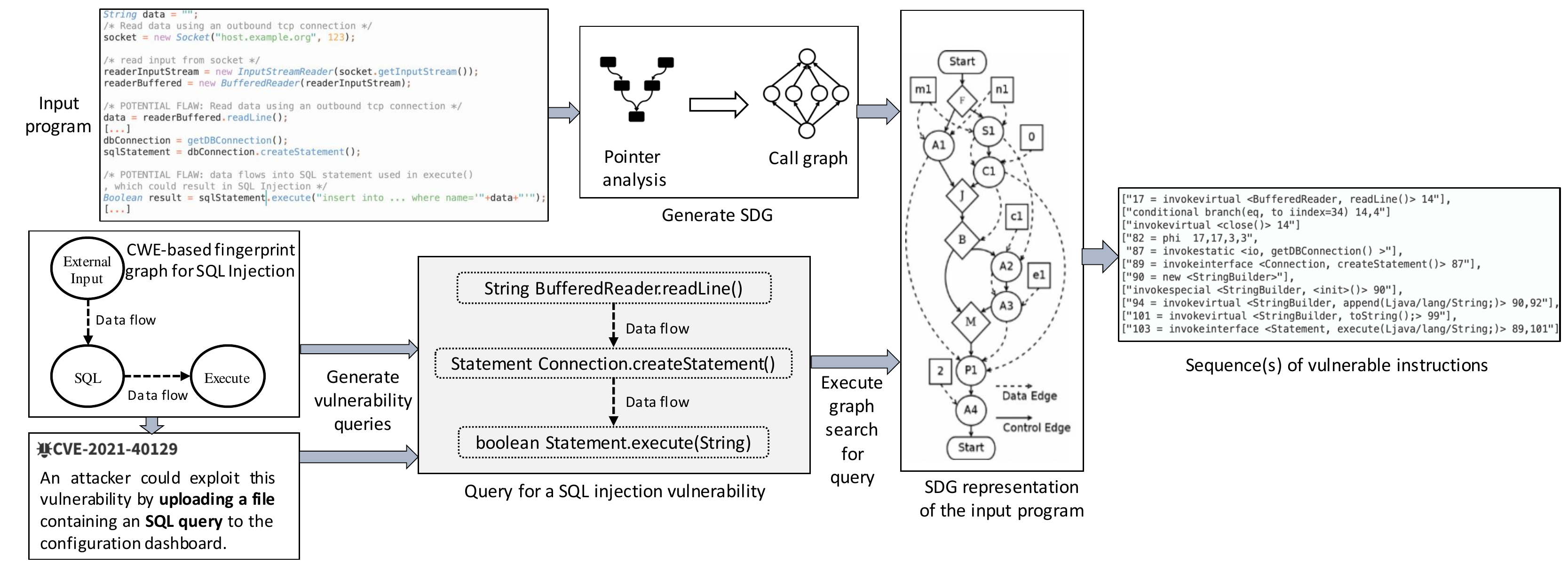}
    \caption{Extracting instruction sequences from CVE and CWE databases and source codes of exploits.}
    \label{fig:CVE2seq}
\end{figure*}

This section describes the extraction of exploit instruction sequences from the CVE and CWE 
databases. We give an overview of this extraction procedure in Fig.~\ref{fig:CVE2seq}. 
First, we use the CWE database to define a high-level vulnerability fingerprint. Although these 
fingerprints can represent the exploit's high-level behavior, they are not sufficient for capturing the 
fine-grained relationships between the program entities (e.g., language-specific APIs). Therefore, we 
use the CVE database to extract the low-level program entities and their relationships that trigger the 
corresponding vulnerabilities. We refer to this low-level representation of the vulnerability fingerprint as 
a vulnerability query. Next, we create the vulnerable program's SDG representation. Finally, we perform 
graph search based on the extracted query to obtain the vulnerable program flow.

\subsubsection{CWE and CVE databases}
We use both CWE and CVE databases to analyze vulnerabilities. The CWE database contains broad categories of 
system vulnerabilities and their high-level descriptions. For example, SQL injection vulnerabilities and cross-site request forgery have been assigned \textit{CWE-89} and \textit{CWE-352}, respectively. 
Therefore, we use the CWE descriptions to extract the high-level semantics of the vulnerabilities in our threat 
model. However, the CWE database does not provide the low-level system-specific vulnerability information that 
we require for real-world exploit classification.
The CVE database contains this low-level information. It contains the specific flaws and weaknesses of 
various systems. Therefore, we obtain the system-specific vulnerability information from the CVE database. 

For example, SQL injection vulnerability has been assigned a CWE, namely CWE-89. However, SQL injection (\textit{CWE-89}) has multiple entries in the CVE database. Each entry corresponds to a unique, vulnerable software version. Some popular CVEs for SQL injection are \textit{CVE-2007-6602} and \textit{CVE-2017-11508}. We utilize the high-level vulnerability semantics from the CWE database and the low-level vulnerability details from the CVE database for further analysis. 


\subsubsection{Vulnerability fingerprints}
We describe the process of extracting vulnerability fingerprints from the CWE database in this section. 
First, we analyze the CWE database to define the high-level vulnerability
semantics. Then, we represent this information with the help of graphs. The nodes of these graphs 
represent entities involved in a particular vulnerability. The edges represent the relationships between these 
entities. These relationships must hold for the vulnerability to be triggered. For example, we can 
describe the semantics of SQL injection (CWE-89) using a fingerprint graph, as shown in Fig.~\ref{fig:CVE2seq}, where 
a vulnerability is triggered when an SQL command is constructed using user-controlled (external) input and executed 
by the system.

\subsubsection{Vulnerability queries}
Vulnerability fingerprints defined by high-level semantics are insufficient for capturing the fine-grained and diverse behaviors of vulnerable programs. For example, user-controllable external inputs may trigger a vulnerability (e.g., SQL injection) in databases. However, user-controllable information can originate from various sources, such as files or network streams. Similarly, different programs might use other APIs for executing SQL commands. To capture this low-level information, we transform the vulnerability fingerprints into vulnerability queries. To construct the queries, we further analyze the CVEs of the corresponding CWE. We extract the relevant entities and their relationships that exploit a vulnerability from the CVE database. Then, we map the entities to their related programming language-specific elements. For example, as shown in Fig.~\ref{fig:CVE2seq}, 
we map a user-controlled external input from files to the corresponding APIs (e.g., \textit{String BufferedReader.readLine()}) for reading the input file. This query representation allows us to check for vulnerable sequences in the SDG representation of the program.

\subsubsection{Vulnerable instruction sequences}
A vulnerable sequence is a series of instructions that exploits a vulnerability. In this section, we describe the process of generating such sequences. As illustrated in Fig.~\ref{fig:CVE2seq}, first, we perform pointer analysis and obtain the precise call graph for an input program. The call graph contains the calling relationships between functions in the program. Then, we process the call graph to obtain the program's SDG. The SDG captures both control- and data-flow 
dependencies of a program. 

Finally, we search the SDG for the vulnerability queries that we generated before. This graph search outputs the sequence of basic blocks and instructions representing the program segment that triggers the vulnerability. We can do this analysis for vulnerability exploits in any programming language. However, we have analyzed Java-based exploits with the Wala code disassembler for our experiments. 

In this work, we use the SDG to extract vulnerable API calls from the programs in our dataset. Prior research
utilizes other graph representations, along with the SDG, to detect vulnerabilities and explore various properties
of the program. For example, SSLint~\cite{he2015vetting} uses a Program Dependence Graph
(PDG) 
that combines both Control Flow Graphs (CFGs) and Data Flow Graphs to
perform intra-procedural analysis for detecting vulnerable properties within a function in an SSL program.
Yamaguchi et al.~\cite{yamaguchi2014modeling} combine an Abstract Syntax Tree (AST), CFG, and PDG to introduce 
a new graph representation called Code Property Graph to analyze lower-level code properties, such 
as \textit{expressions} and \textit{conditions} in a program. While all these representations have proven to be 
effective for various applications, the SDG representation is the best candidate for our problem. An SDG, being an 
inter-procedural representation of the PDG, combines the control flow and data flow relationships between the APIs in a program. This inter-procedural representation of vulnerable programs enables us to capture the holistic program execution environment.

However, SDG construction can be imprecise for certain program features (e.g., reflected function calls) and can generate sequences that may not occur at run-time in the given program. Fortunately, this is an advantage for the generality of our training model. In other words, since an imprecise (i.e., overapproximated) SDG for certain program components can represent a program path that may not occur at run-time, it allows our model to learn those paths as well, since they could occur at run-time in other programs with different inputs.


\subsection{Feature Engineering}
\label{sec:Feature_engineering}
We need to convert the instruction (or API) calls of execution traces into feature vectors that can be analyzed by 
ML models. Unfortunately, most prior research ignores relevant information in the API calls during 
feature extraction. Some omitted discriminative information includes API call arguments, API output 
types, API categories, and API analysis scope. Table~\ref{tab:comparison_chart} compares the feature representation of ML-FEED with prior research. 

\begin {table}[h]
\vspace{0.5cm}
\centering
\caption {Feature representation comparison} 
\label{tab:comparison_chart}
\scalebox{1}{
\begin{tabular}{|c|c|c|c|c|c|}
\hline
\textbf{Model} & \textbf{1} & \textbf{2} & \textbf{3} & \textbf{4} & \textbf{5} \\ \hline

Roy et al. \cite{roy2020android} & \xmark & \cmark & \xmark & \xmark & \cmark \\ \hline
Russel et al. \cite{russell2018automated}& \cmark & \xmark & \xmark & \cmark & \cmark \\ \hline
MaMaDroid \cite{onwuzurike2019mamadroid} & \cmark & \cmark & \xmark & \cmark & \cmark \\ \hline
VulDeePecker \cite{li2018vuldeepecker} & \cmark & \xmark & \xmark & \xmark & \cmark \\ \hline
$\mu$-VulDeePecker \cite{zou2019muvuldeepecker} & \cmark & \xmark & \xmark & \xmark & \cmark \\ \hline
Yamaguchi et al. \cite{yamaguchi2011vulnerability} & \cmark & \xmark & \xmark & \cmark & \cmark \\ \hline
SHARKS \cite{saha2021sharks} & \xmark & \cmark & \xmark & \cmark & \xmark \\ \hline
GRAVITAS \cite{brown2021gravitas} & \xmark & \cmark & \xmark & \cmark & \xmark \\ \hline
\textbf{ML-FEED} & \cmark & \cmark  & \cmark & \cmark & \cmark \\ \hline

\end{tabular}}
\end {table}

The various columns in Table \ref{tab:comparison_chart} are as follows: (1) API name, (2) API category, (3) API analysis scope, (4) API package and class, and (5) API input/output type.

There are two reasons for the omission of various features in prior research. First, converting all 
the information into vectors is often a non-trivial task. Second, more information in the feature vector 
increases vector size. Higher-dimensional feature vectors require more sophisticated ML models. However, including this information 
leads to better performance~\cite{roy2020android}. For example, ignoring the API call parameters for a read operation
can make the feature vectors of all read operation-based API calls similar. However, the read operation 
may be malicious if the input comes from the user, whereas it is benign if it is system-generated. 
Therefore, we should not ignore this information in order to avoid misrepresenting API calls. In our 
experiments, we do not overlook any such information. Table \ref{tab:Features} shows the details of the 
feature representations that we use in ML-FEED.
It uses a hybrid feature engineering approach that combines the benefits of NLP embeddings, one-hot 
encoding, and frequency vectors.

\begin {table}[]
\centering
\vspace{0.4cm}
\caption {An overview of feature representation}
\begin{tabular}{|p{1.6cm}|p{1.2cm}|>{\centering\arraybackslash}p{1.2cm}|p{2cm}|}
\hline
\textbf{Feature name} & \textbf{Feature encoding} & \textbf{Dimension} & \textbf{Examples}\\ \hline
API name & NLP & 70 & readLine()
\\ \hline
API category & One-hot & 9 & invokespecial
\\ \hline
API analysis scope & One-hot & 2 & Application
\\ \hline
API package & One-hot & 22 & Ljava/io/ BufferedReader
\\ \hline
API input/ output type & Frequency vector & 24 & Ljava/sql/ ResultSet
\\ \hline
\end{tabular}
\label{tab:Features}		
\end {table}

\subsubsection{NLP embeddings}
One of the many challenges of vulnerability exploit detection is detecting semantically-similar APIs. 
For example, if an  API $f_1$ triggers an exploit, then probably, a semantically-identical process 
$f_2$ may also trigger the same exploit. Hence, it is essential to capture the behavioral semantics of 
the APIs. We observe that the API names are good indicators of behavioral semantics. For example, the 
APIs for opening a new file in Java are \textit{newFile()} and \textit{createFile()}. This observation 
inspires us to use NLP-based embeddings to encode the API names.

We use a pre-trained lightweight word2vec embedding model, called \textit{FastText}~\cite{joulin2016fasttext}, to compute the embedding of 
the API name. Word2vec is an NLP model that learns word associations from a large text corpus. A trained 
word2vec model is capable of detecting synonymous words. It outputs vector representations of words such 
that similar words are closer in the vector space. These vector representations of words are known as 
word embeddings. FastText is trained on the text corpus of Wikipedia and returns 10-dimensional word 
embeddings. In our dataset, no API name has more than seven words. Hence, we concatenate the seven 
embeddings to obtain a 70-dimensional feature vector for the API name. If an API name has less than seven 
words, we fill in the missing values with zeros.


\begin{figure}[h]
    \centering
    \includegraphics[scale=0.35]{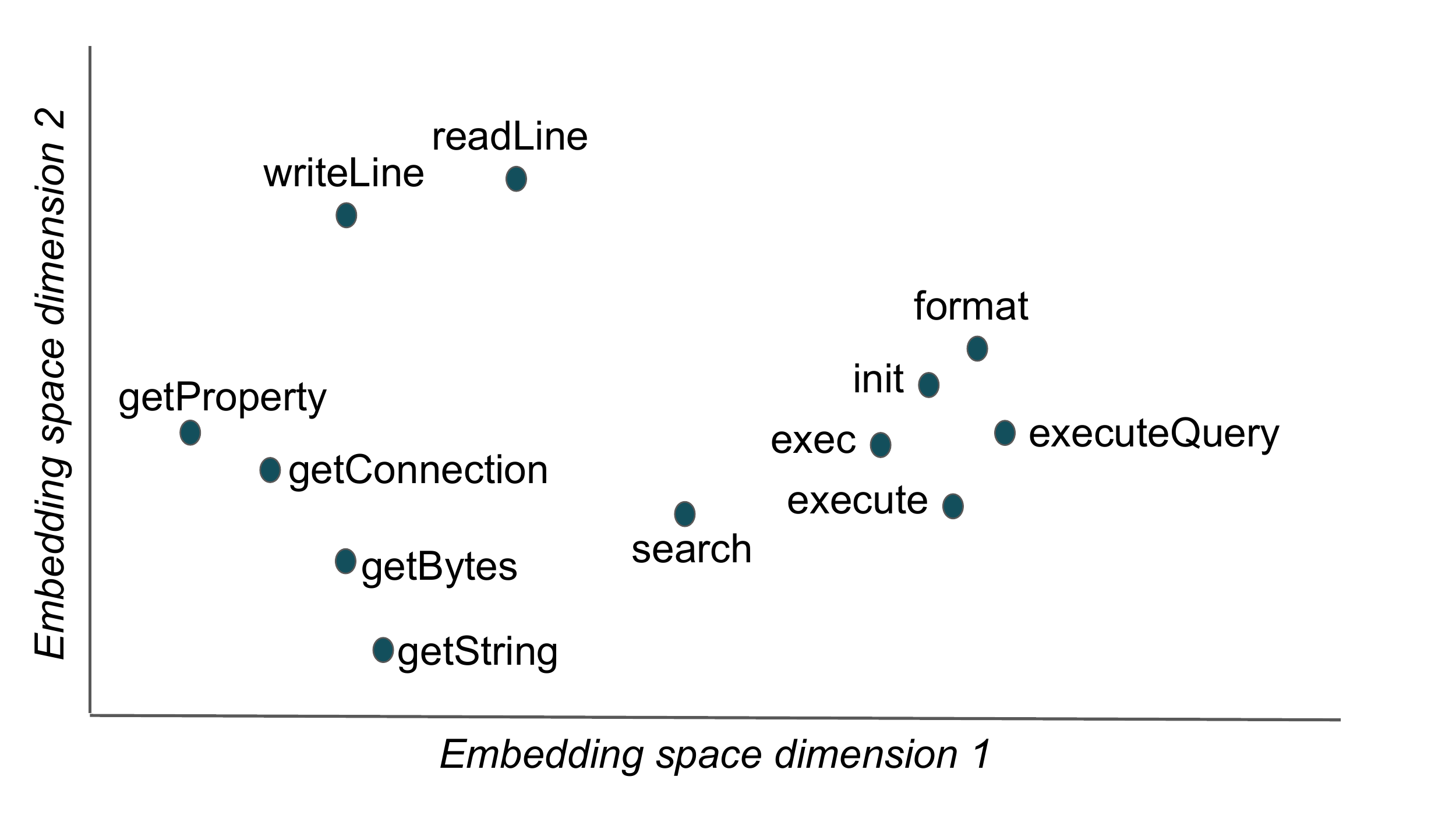}
    \caption{A plot showing the NLP embeddings of some of the APIs in 2D. The 70-dimensional embedding 
has been reduced to 2D using PCA for visualization.}
    \label{fig:embeddings}
    
\end{figure}

\begin{figure*}[h]
    \centering
    \includegraphics[scale=0.18]{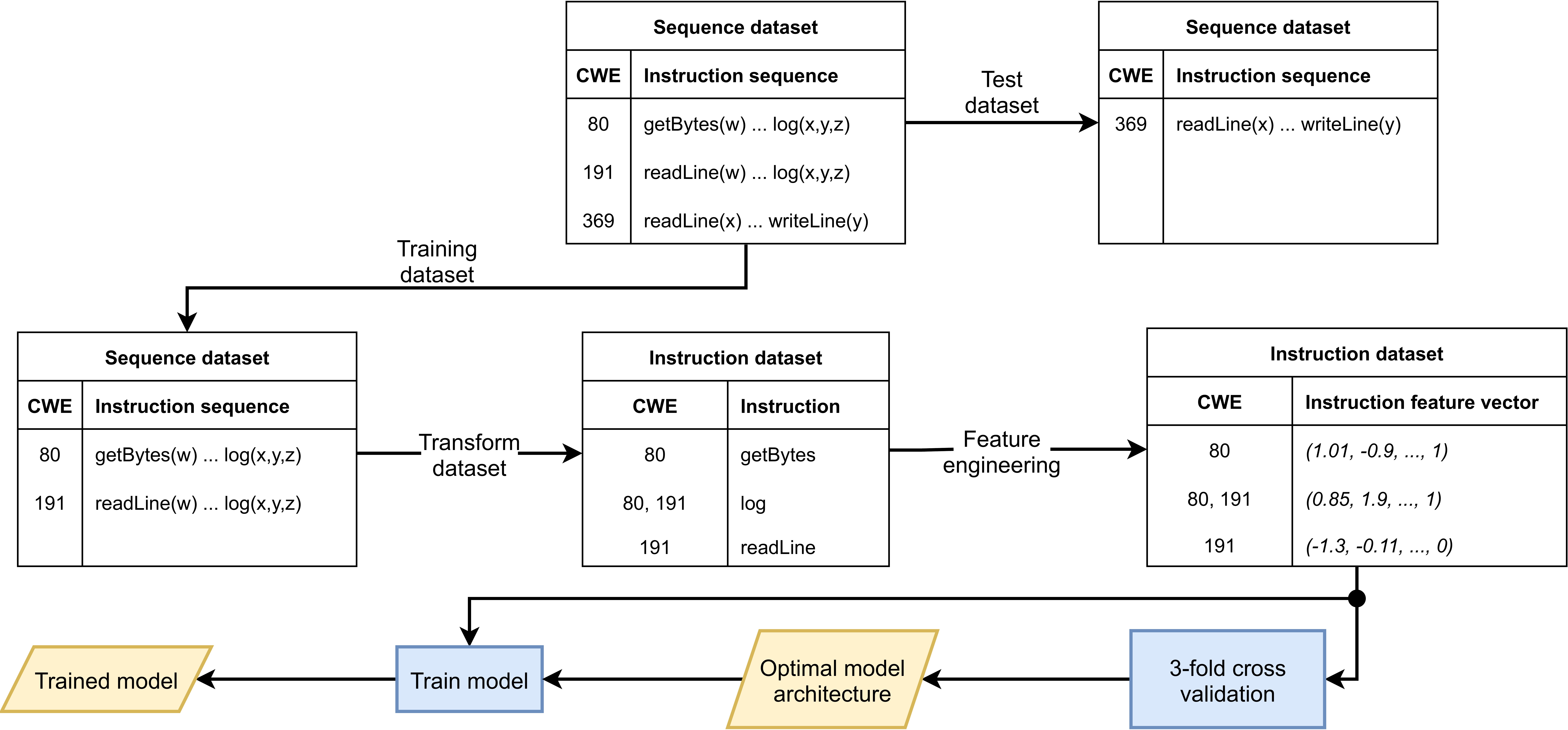}
    \caption{Overview of design and training of the ML-FEED exploit classifier}
    \label{fig:training-pipeline}
\end{figure*}


\subsubsection{One-hot encoding}
We have many categorical features in our dataset. For example, the API analysis scope has two 
categories: Application and Primordial. We experimented with label encoding and one-hot encoding to 
encode the categorical features.

Label encoding assigns a unique integer value to every possible feature category. Unfortunately, 
it suffers from spurious correlations. To avoid this problem, we use the 
one-hot encoding method. For categorical features with $n$ possible types, a one-hot encoder creates 
an $n$-dimensional binary vector. Every category is assigned a unique index in the vector. For example, 
to encode category $i$ via one-hot encoding, only the $i^{th}$ element of the vector is assigned a 1 and the 
remaining elements are assigned 0's. We demonstrate label encoding and one-hot encoding of all the categories of API types in Table \ref{tab:Encoding}.


\begin {table}[h]
\centering
\vspace{1mm}
\caption {Example feature encodings of API types}
\begin{tabular}
{| >{\centering\arraybackslash} p{1.45cm}| >{\centering\arraybackslash} p{1.2cm}|c|c|c|c|c|c|c|c|c|}
\hline
\textbf{Feature} & \textbf{Label encoding} & \multicolumn{9}{c|}{\textbf{One-hot encoding}}\\ \hline
binaryop & 1 & 1 & 0 & 0 & 0 & 0 & 0 & 0 & 0 & 0\\ \hline
conversion & 2 & 0 & 1 & 0 & 0 & 0 & 0 & 0 & 0 & 0\\ \hline
getCaught Exception & 3 & 0 & 0 & 1 & 0 & 0 & 0 & 0 & 0 & 0\\ \hline
getstatic & 4 & 0 & 0 & 0 & 1 & 0 & 0 & 0 & 0 & 0\\ \hline
invoke interface & 5 & 0 & 0 & 0 & 0 & 1 & 0 & 0 & 0 & 0\\ \hline
invoke special & 6 & 0 & 0 & 0 & 0 & 0 & 1 & 0 & 0 & 0\\ \hline
invoke static & 7 & 0 & 0 & 0 & 0 & 0 & 0 & 1 & 0 & 0\\ \hline
invoke virtual & 8 & 0 & 0 & 0 & 0 & 0 & 0 & 0 & 1 & 0\\ \hline
phi & 9 & 0 & 0 & 0 & 0 & 0 & 0 & 0 & 0 & 1\\ \hline
\end{tabular}
\label{tab:Encoding}		
\end {table}


\subsubsection{Frequency vector}
Frequency vectors contain the frequencies of various elements in the API call. We use them to encode the API 
input/output (I/O) types, which denote the frequency of every I/O class among the API call parameters. We present an example of frequency vectors in Table \ref{tab:Frequency_vector}. For conciseness, we only show 4-dimensional frequency vectors in the table. 
As shown in Table \ref{tab:Features}, we have 24 unique API input/output types. Therefore, we have a 24-dimensional frequency vector representing the API call input types. Likewise, we have another 24-dimensional frequency vector representing the API output types. 


\begin {table}[h]
\centering
\vspace{0.5cm}
\caption {Example frequency vector-based encodings of API call parameters of Java APIs}
\begin{tabular}
{|c|c|c|c|c|}
\hline
\textbf{API name} & \multicolumn{4}{c|}{\textbf{Input/Output types}}\\ \hline
& \textit{String} & \textit{Level} & \textit{Throwable} & \textit{File}\\ \hline
log & 1 & 1 & 1 & 0 \\ \hline
getString & 1 & 0 & 0 & 0 \\ \hline
\end{tabular}

\label{tab:Frequency_vector}		
\end {table}


\subsubsection{Combination of encodings}
We concatenate the hybrid encodings of various elements to construct the final feature vector for an instruction call. 
We show the dimension of every feature in Table~\ref{tab:Features}. We have two feature vectors corresponding to 
the last row in the table -- one for the API inputs and another for the API outputs. We have one 
feature vector for all the other rows. Thus, the concatenated feature vector of the instruction call has a dimension of 151.

Next, we describe how our model, ML-FEED, classifies exploits efficiently. ML-FEED is a cascade of a classifier and a dynamic state table. Unlike most prior research, ML-FEED does not analyze entire sequences of instruction calls at once. It looks into the problem at a finer granularity and analyzes one instruction call at a time. This strategy eliminates the necessity of using sophisticated sequence models, resulting in reduced computation overhead. The ML-FEED classifier module takes the 151-dimensional feature vector of an instruction call as input and outputs the potential exploits that the instruction call might trigger. 

We show the pipeline for training the ML-FEED exploit classifier in Fig.~\ref{fig:training-pipeline}. The training procedure consists of the following steps. First, we split the dataset of vulnerable instruction sequences into a training set with 85\% of the data and a test set with the remaining 15\%. We use the training set to extract individual instructions from the sequences and label them with the exploits that they can potentially execute. Then, we convert these instructions into labeled feature vectors. We describe the process of obtaining such vectors from the dataset of instruction sequences in Section~\ref{sec:dataset-transformation}. Next, we use the dataset of labeled feature vectors to design the neural network for our classifier. We use 3-fold cross-validation to select the optimal design of our neural network. We elaborate on this process in Section~\ref{sec:classifier-model-architecture}. Finally, we train the optimal neural network architecture with the training dataset to get our final model.

\subsubsection{Transforming the dataset} 
\label{sec:dataset-transformation}
As mentioned in Section \ref{sec:Representation_vuln_programs}, we convert the 
exploits in the CWE database to vulnerable instruction sequences. We need to extract the features and
labels of the individual instruction calls from the dataset of instruction sequences. We describe this
information extraction procedure next. First, we obtain the feature vectors of all the instruction calls
via the feature engineering method discussed in Section \ref{sec:Feature_engineering}. Then, we generate
the labels for all the feature vectors. The labels contain exploits that the corresponding instruction
calls can trigger. A single instruction call can trigger multiple exploits.  The exploits are generally
independent of each other. Thus, we need a multi-label classifier that predicts all the exploits that a
given instruction can execute. We design our classifier output to be an $n$-dimensional binary vector to
enable this. We assign the values corresponding to the triggered exploits in this vector as $1$, whereas
the rest are assigned $0$. For example, if instruction call \textit{A} executes exploits 1 and 23, only the first and $23^{rd}$ elements of the $n$-dimensional output vector will be $1$. 

\subsubsection{Classifier model architecture} 
\label{sec:classifier-model-architecture}
The classifier is a fully-connected feedforward neural network. We 
need to design a lightweight neural network that achieves high accuracy. The neural network architecture has the following 
constraint -- the number of neurons in the input and output layers is pre-determined by the input and output 
dimensions. Therefore, we only need to determine the number of hidden layers and the number of neurons in each 
hidden layer. Furthermore, we target a lightweight model to reduce its latency. Hence, we start with one hidden layer. 
We observe that architectures with one hidden layer do not produce good results. 
Thus, we expand our search to two hidden layers. Two hidden layers achieve 
much better results. Finally, we reduce the number of neurons in the hidden layers to achieve the most lightweight 
model with comparable accuracy. 
Table~\ref{tab:Classifier} describes our chosen 
architecture. The output layer has 79 neurons because the output of the classifier is a 79-dimensional binary vector that denotes which of the 79 exploit categories are executed at a given time.

\begin {table}[h]
\centering
\vspace{0.35cm}
\caption {The model architecture of the ML-FEED classifier}
\begin{tabular}{|c|c|}
\hline
 & \textbf{Number of neurons}\\ \hline
Input layer & 151\\ \hline
Hidden layer 1 & 150\\ \hline
Hidden layer 2 & 100\\ \hline
Output layer & 79\\ \hline
\end{tabular}
\label{tab:Classifier}
\end {table}

We train this classifier model on the complete training dataset and evaluate it on the test dataset.

\subsubsection{Sequence monitor}
\label{sec:sequence-monitor}
Cyber-attackers can exploit a vulnerability by executing a set of malicious instructions in a particular sequence. 
However, running the same set of instructions in a different order may not trigger the vulnerability. Thus, we 
need to track the series of program statements and instruction calls to detect exploits. Traditional sequence models, 
namely LSTMs and transformers, can track the order implicitly. However, this leads to high computation overheads. 
Therefore, we propose a state table-based approach to track sequences of program instructions.

First, we discuss a naive pattern matching approach that can be implemented with a state table and its drawbacks. Then, we show how we can overcome these drawbacks with ML-FEED. Every row in the state 
table corresponds to a unique exploit. We can initialize the exploit states with the first instruction call in the 
respective exploit fingerprints in the naive approach. When we encounter any instruction call in the current state table 
during program execution, we update the corresponding states with the following instruction call of the exploit 
fingerprint. If the encountered instruction call was the last one in the fingerprint(s), we would raise an alarm 
indicating that the API call has executed the corresponding exploit(s). Since we analyze the API call parameters 
like the API name, input types, and API package name, this method is also applicable to obfuscated codebases. This naive approach is similar to the subgraph matching approaches used for vulnerability detection~\cite{aaraj2008dynamic}.

Although the naive method is much more lightweight than sequence models, it has two drawbacks that limit its 
applicability in real-world scenarios. First, the naive pattern matching approach cannot detect 
semantically-similar instruction calls. As a result, the attacker may avoid detection by executing different APIs with 
the same functionality as the instruction in the fingerprint. Second, checking the existence of every instruction 
in the state table is a time-intensive task. This timing overhead reduces the usefulness of naive pattern matching 
in real-time scenarios~\cite{aaraj2008dynamic}. 

Our solution, ML-FEED, addresses these drawbacks of naive pattern matching. Instead of storing the instructions 
in the state table, we store the feature vectors of the instruction calls. Then, we compare the feature vectors of the executed APIs with the entries in the state table. We use cosine similarity to measure the similarity between 
them. This method enables us to capture semantic similarities between APIs. If the cosine similarity is above a 
threshold value, we update the state table with the feature vector of the following API in the exploit chain. We 
experimentally observed that a threshold value of 0.9 enables the model to achieve the highest F1 score and accuracy.

To address the second drawback of considerable time overheads, we do not compare the feature vector of the executed API 
with all the states in the state table. Instead, we compare it with only the states of the exploit categories 
predicted by the ML-FEED classifier. 
Our experiments show that this method requires nearly $10\times$ fewer comparisons per instruction call.

\subsubsection{Instruction white-list}
To further reduce inference time overheads, ML-FEED also uses a pre-defined white-list of instructions that are not relevant to the sensitive API calls. The instructions in the white-list are those that do not have any control or data dependencies on the exploit chain~\cite{aaraj2008dynamic}. For example, Java uses the \textit{getCaughtException} API to handle exception calls during program execution. The presence of this API call has no impact on the success or failure of triggering an exploit in our threat model. Therefore, 
We can safely discard frequently used white-listed APIs without losing sensitive data and control dependency information. Furthermore, this ignorance of non-sensitive API calls does not affect ML-FEED's ability to detect potential exploits.

\subsubsection{Inference pipeline}
\begin{figure*}[t]
    \centering
    \includegraphics[scale=0.16]{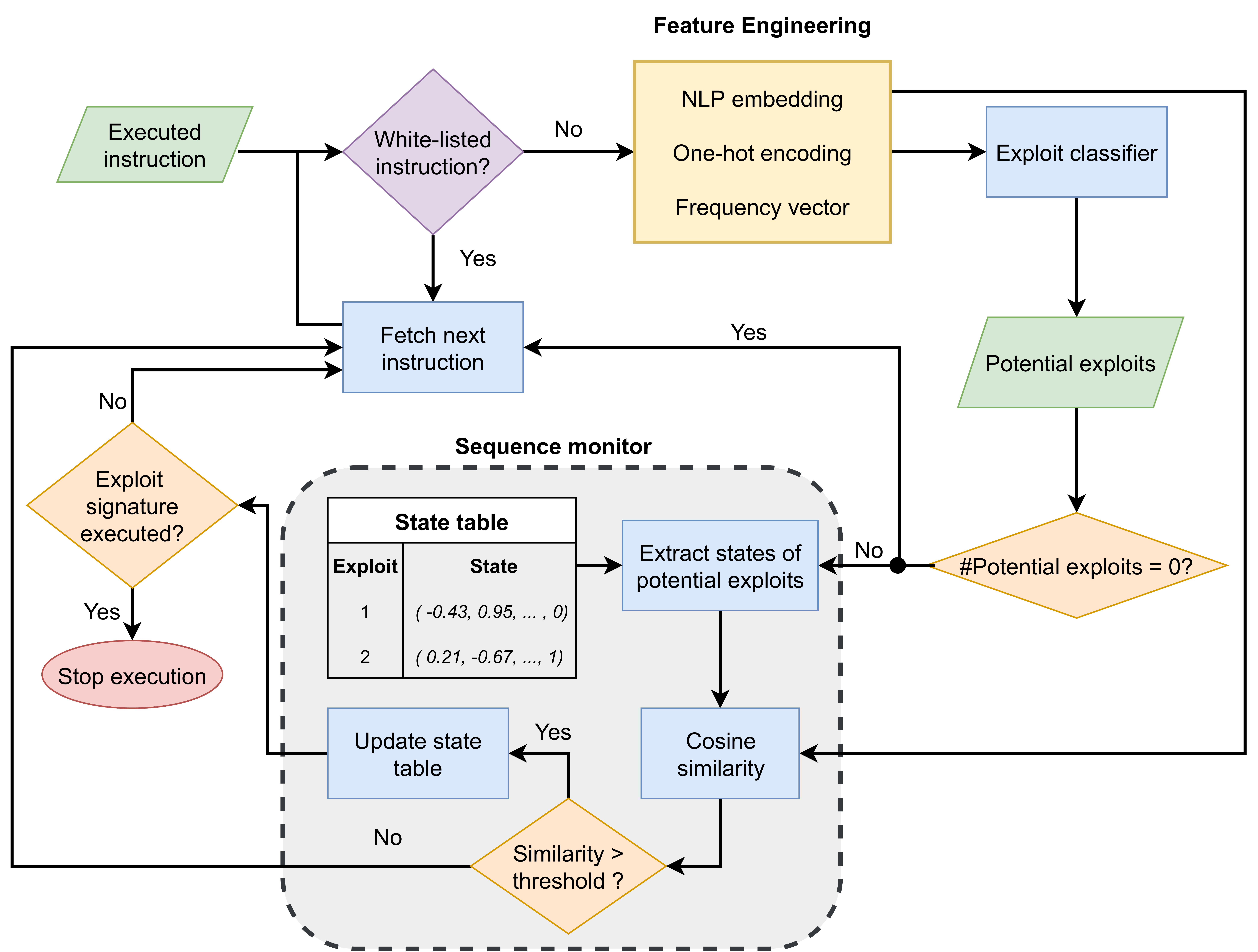}
    \caption{The ML-FEED inference pipeline}
    \label{fig:inference-pipeline}
\end{figure*}

We have designed ML-FEED to efficiently detect exploits without compromising accuracy. 
We demonstrate our end-to-end inference pipeline in Fig.~\ref{fig:inference-pipeline}. As shown in the figure, we input the executed instruction into our framework at run-time. We proceed to the next executed instruction if the executed instruction is present in our white-list of non-sensitive API calls. Otherwise, we analyze the instruction to determine if it triggers an exploit. First, we obtain the feature vector for the executed instruction. The ML-FEED exploit classifier processes this feature vector. Next, the classifier outputs the list of potential exploits that the executed instruction might trigger. If this list is empty, we terminate our analysis of the current instruction and proceed to the next executed instruction. Otherwise, we input the list of potential exploits to the ML-FEED sequence monitor. The sequence monitor has a dynamic state table that stores the states of all the exploits in our threat model. From the state table, we extract the exploit states of all the potential exploits predicted by the classifier. Next, we compute the cosine similarities between the states of the extracted exploits with the feature vector of the executed instruction. As mentioned in Section~\ref{sec:sequence-monitor}, we can successfully detect semantically-similar instructions by making this comparison. 
Finally, we update the exploit states similar to the feature vector of the executed instruction with the feature vectors of the following instruction in the corresponding exploit fingerprints. However, if the state table indicates that the executed instruction has implemented a potential vulnerability fingerprint, we stop the execution and raise an alarm. Otherwise, we proceed to analyze the next instruction.

\section{Experimental Setup and Evaluation}
\label{sec:eval}

This section describes our experimental setup, provides implementation details, and discusses evaluation results.

\subsection{Experimental Setup}
In what follows, we discuss our vulnerability database construction and the experimental setup for the evaluation.

\subsubsection{Vulnerability database}
We train our model with 23 unique CWEs from the CWE dataset. Table \ref{tab:CWE} shows the complete 
list of the CWE vulnerabilities that we consider. As shown in the table, we evaluate multiple exploits and payloads 
for each CWE, resulting in an analysis of 79 exploits. Based on the vulnerability database, we construct an exploit database from the Juliet Test Suite~\cite{juliet} for Java, containing code examples for 112 CWEs. 
Given a code example from this suite, we first construct the Java bytecode using \textit{the javac} 
compiler. We then use Wala to perform pointer analysis on the bytecode and build the call graphs. 
Finally, we create the SDG using pointer analysis and call graphs. A node in the SDG represents an instruction 
and an edge represents control or data flow between instructions. Although our method can analyze exploits in any 
programming language, we focus on Java-based exploits for our evaluation.

\begin {table}[h]
\centering
\vspace{0.5cm}
\caption {Details of CWEs considered under our threat model}
\begin{tabular}
{|c| >{\centering\arraybackslash} p{4.5cm}| >{\centering\arraybackslash} p{1.5cm}|}
\hline
\textbf{CWE id} & \textbf{Description} & \textbf{Number of exploits}\\ \hline
CWE-15 & External control of system or configuration settings & 2\\ \hline
CWE-23 & Relative path traversal & 1\\ \hline
CWE-78 & OS command injection & 2\\ \hline
CWE-81 & Cross-site scripting & 1\\ \hline
CWE-89 & SQL injection & 10\\ \hline
CWE-90 & LDAP injection & 1\\ \hline
CWE-129 & Improper validation of array index & 14\\ \hline
CWE-134 & Uncontrolled format string & 3\\ \hline
CWE-190 & Integer overflow & 3\\ \hline
CWE-191 & Integer underflow & 5\\ \hline
CWE-197 & Numeric truncation error & 6\\ \hline
CWE-226 & Sensitive information uncleared before release & 1\\ \hline
CWE-256 & Plaintext storage of password & 1\\ \hline
CWE-259 & Hard-coded password & 1\\ \hline
CWE-319 & Sensitive information in clear text & 2\\ \hline
CWE-369 & Division by zero & 10\\ \hline
CWE-400 & Resource exhaustion & 6\\ \hline
CWE-470 & Unsafe reflection & 1\\ \hline
CWE-506 & Embedded malicious code & 1\\ \hline
CWE-570 & Expression always false & 1\\ \hline
CWE-606 & Unchecked loop condition & 3\\ \hline
CWE-643 & Xpath injection & 1\\ \hline
CWE-789 & Uncontrolled memory allocation & 3\\ \hline

\end{tabular}

\vspace{-0.2cm}

\label{tab:CWE}		
\end {table}

We represent every exploit with code snippets in order to cover all the features of a CWE.  We generate 226,207 code snippets to represent 79 exploits across 23 CWEs. We also generate 1,101,388 benign code snippets to represent non-vulnerable code executions. Thus, the total size of our database is 1,327,595 instruction sequences. We split these sequences into training and test set with a train-test split of 85\%-15\%.

\subsubsection{Exploit database}

We construct our exploit database from the Juliet Test Suite~\cite{juliet} for Java, containing code examples for 112 CWEs. Given a code example from this suite, we first construct the Java bytecode using \textit{the javac} compiler. We then use Wala to perform pointer analysis on the bytecode and build the call graphs. 

Finally, we create the SDG using pointer analysis and call graphs. A node in the SDG represents an instruction, and an edge represents control or data flow between instructions. Although our method can analyze exploits in any programming language, we focus on Java-based exploits for our evaluation. 

\subsubsection{Parameters}

We use the most lightweight state-of-the-art LSTM and transformer models for comparison to ML-FEED. The hidden state of the LSTM cell is a 6-dimensional vector. The lightweight GPT-2 transfomer that we use in our experiments has 117 million parameters and 12 decoder layers with 12 attention heads in each self-attention layer. The GPT-2 model processes a series of 1024 API calls at any given time instance.

\subsubsection{Machine configuration}
We ran experiments on a Lenovo ThinkPad laptop with an Intel Core i7-8665U CPU. The CPU has four cores with a 
maximum operating frequency of 4.2 GHz for each core. The processor base frequency is 1.90GHz.


\begin {table*}[]
\centering
\caption {Performance of ML models. All values depict percentages (\%). The best performances are denoted in \textbf{bold}.}
\begin{tabular}{|c|P{1.5cm}|cccc|cccc|cccc|}
\hline
& & \multicolumn{4}{c|}{\textbf{LSTM}} &
\multicolumn{4}{c|}{\textbf{Transformer}} & 
\multicolumn{4}{c|}{\textbf{ML-Feed}} \\ 
\hline
\textbf{CWE id} & \textbf{OWASP top-10} & \textit{Acc.} & \textit{Prec.} & \textit{Rec.} & \textit{F1} & \textit{Acc.} & \textit{Prec.} & \textit{Rec.} & \textit{F1} & \textit{Acc.} & \textit{Prec.} & \textit{Rec.} & \textit{F1} \\
\hline
CWE-15 & \cmark & 99.5 & 93.8 & 98.9 & 96.2 & 99.6 & 96.3 & 99.0 & 97.6 & \textbf{99.8} & \textbf{97.5} & \textbf{99.6} & \textbf{98.6}\\ 
\hline
CWE-23 & \cmark & \textbf{96.4} & 93.6 & 94.9 & \textbf{96.4} & 93.6 & 94.9 & \textbf{96.4} & 93.6 & 94.9 & \textbf{96.4} & 93.6 & 94.9\\ 
\hline
CWE-78 & \cmark & \textbf{99.9} & 91.2 & 73.8 & 81.6 & \textbf{99.9} & 91.4 & 76.2 & 83.1 & \textbf{99.9} & \textbf{91.9} & \textbf{80.9} & \textbf{86.0}\\ 
\hline
CWE-81 & \xmark & \textbf{99.9} & 91.9 & 80.9 & 86.0 & \textbf{99.9} & 92.1 & 83.3 & 87.5 & \textbf{99.9} & \textbf{92.5} & \textbf{88.1} & \textbf{90.2}\\ 
\hline
CWE-89 & \cmark & 99.2 & 93.8 & 99.2 & 96.4 & 99.5 & 96.3 & 99.4 & 97.8 & \textbf{99.7} & \textbf{97.5} & \textbf{99.7} & \textbf{98.6}\\ 
\hline
CWE-90 & \cmark & 99.1 & 92.9 & 69.0 & 79.2 & \textbf{99.9} & 95.2 & 72.7 & 82.5 & \textbf{99.9} & \textbf{96.4} & \textbf{86.2} & \textbf{91.0}\\ 
\hline
CWE-129 & \xmark & 99.8 & 93.8 & 93.9 & 93.9 & \textbf{99.9} & 96.2 & 96.2 & 96.2 & \textbf{99.9} & \textbf{97.4} & \textbf{98.0} & \textbf{97.7}\\ 
\hline
CWE-134 & \xmark & \textbf{99.9} & 93.3 & 52.5 & 67.2 & \textbf{99.9} & \textbf{95.6} & 61.4 & 74.8 & \textbf{99.9} & \textbf{95.6} & \textbf{72.9} & \textbf{82.7}\\ 
\hline
CWE-190 & \xmark & 99.5 & 93.8 & 98.9 & 96.3 & 99.7 & 96.2 & 99.3 & 97.7 & \textbf{99.8} & \textbf{97.6} & \textbf{99.7} & \textbf{98.6}\\ 
\hline
CWE-191 & \xmark & \textbf{99.9} & 93.2 & 79.9 & 86.1 & \textbf{99.9} & 95.8 & 92.5 & 94.1 & \textbf{99.9} & \textbf{97.4} & \textbf{93.0} & \textbf{95.2}\\ 
\hline
CWE-197 & \xmark & \textbf{99.9} & 92.6 & 41.7 & 57.5 & \textbf{99.9} & 92.6 & 50.0 & 64.9 & \textbf{99.9} & \textbf{96.3} & \textbf{74.3} & \textbf{83.9}\\ 
\hline
CWE-226 & \xmark & \textbf{99.9} & 93.4 & 85.4 & 89.2 & \textbf{99.9} & 96.0 & 91.9 & 93.9 & \textbf{99.9} & \textbf{97.3} & \textbf{95.6} & \textbf{96.5}\\ 
\hline
CWE-256 & \cmark & \textbf{99.9} & 93.4 & 85.4 & 89.2 & \textbf{99.9} & 96.0 & 91.9 & 93.9 & \textbf{99.9} & \textbf{97.3} & \textbf{95.6} & \textbf{96.5}\\ 
\hline
CWE-259 & \cmark & \textbf{99.9} & 93.2 & 83.6 & 88.2 & \textbf{99.9} & 95.9 & 82.1 & 88.5 & \textbf{99.9} & \textbf{97.3} & \textbf{92.3} & \textbf{94.7}\\ 
\hline
CWE-319 & \cmark & \textbf{99.9} & \textbf{50.0} & 2.1 & 4.0 & \textbf{99.9} & \textbf{50.0} & 3.8 & 7.1 & \textbf{99.9} & \textbf{50.0} & \textbf{4.3} & \textbf{8.0}\\ 
\hline
CWE-369 & \xmark & 99.6 & 93.8 & 98.7 & 96.2 & 99.7 & 96.2 & 99.2 & 97.7 & \textbf{99.8} & \textbf{97.5} & \textbf{99.5} & \textbf{98.5}\\ 
\hline
CWE-400 & \xmark & \textbf{99.9} & \textbf{88.9} & 20.5 & 33.3 & \textbf{99.9} & \textbf{88.9} & \textbf{32.0} & \textbf{47.1} & \textbf{99.9} & \textbf{88.9} & 26.7 & 41.0\\ 
\hline
CWE-470 & \cmark & \textbf{99.9} & 92.7 & 65.4 & 76.7 & \textbf{99.9} & 94.5 & 72.2 & 81.9 & \textbf{99.9} & \textbf{96.4} & \textbf{80.3} & \textbf{87.6}\\ 
\hline
CWE-506 & \xmark & \textbf{99.9} & 93.5 & 72.5 & 81.6 & \textbf{99.9} & 95.6 & \textbf{88.6} & \textbf{92.0} & \textbf{99.9} & \textbf{97.1} & 85.3 & 90.8\\ 
\hline
CWE-570 & \xmark & \textbf{99.9} & 75.0 & 8.1 & 14.6 & \textbf{99.9} & 95.6 & \textbf{88.6} & \textbf{92.0} & \textbf{99.9} & \textbf{97.1} & 85.3 & 90.8\\ 
\hline
CWE-606 & \xmark & \textbf{99.9} & 85.7 & 75.0 & 80.0 & \textbf{99.9} & 86.8 & 82.5 & 84.6 & \textbf{99.9} & \textbf{87.2} & \textbf{85.0} & \textbf{86.1}\\ 
\hline
CWE-643 & \cmark & \textbf{99.9} & 92.7 & 65.4 & 76.7 & \textbf{99.9} & 94.5 & 72.2 & 81.9 & \textbf{99.9} & \textbf{96.4} & \textbf{80.3} & \textbf{87.6}\\ 
\hline
CWE-789 & \xmark & \textbf{99.9} & 93.6 & 55.1 & 69.4 & \textbf{99.9} & 95.2 & 76.9 & 85.1 & \textbf{99.9} & \textbf{96.8} & \textbf{79.2} & \textbf{87.1}\\ 
\hline
\end{tabular}
\vspace{-0.2cm}
\label{tab:individual_performance}		
\end{table*}

\subsection{Experimental Results}
This section provides experimental results obtained by running ML-FEED and state-of-the-art ML-based sequence models. We report the performance metrics of our model on every CWE in our threat model in Table~\ref{tab:individual_performance}. We also denote the CWEs that belong to the OWASP top-10 vulnerabilities. The non-profit organization, OWASP, ranks the most critical web application security risks based on their frequency of occurrence, severity, and magnitude of the potential impact on the system. Security analysts frequently use the OWASP top 10 vulnerability list as a guideline.

\subsubsection{Metrics}
We define the performance metrics in terms of true positives (TP), true negatives (TN), false positives (FP), and false negatives (FN). TP and TN refer to the number of correctly classified positive and negative labels. FP and FN refer to the number of incorrect positive and negative predictions, respectively. These values are often represented in the form of a confusion matrix. We use these metrics to compute the accuracy, precision, recall, and F1-score of the models.


\begin{enumerate}
    \item \textit{Accuracy:} Accuracy denotes the overall performance of the classifier. Although it is a good 
evaluation metric, it suffers from biased results when evaluating imbalanced datasets.
    
    \begin{equation}
        Accuracy = \frac{TP + TN}{TP + TN + FP + FN} \nonumber
    \end{equation}
    
    \item \textit{Precision:} Precision measures what fraction of the positive predictions of the model is correct. We can compute the precision value from the confusion matrix of a binary classification problem with the following formula.
    
    \begin{equation}
        Precision = \frac{TP}{TP + FP} \nonumber
    \end{equation}
    
    \item \textit{Recall:} Recall measures what fraction of the true positive labels in a binary dataset are predicted as positive. The formula for computing the recall value from the confusion matrix of a binary classification problem is as follows.
    
    \begin{equation}
        Recall = \frac{TP}{TP + FN} \nonumber
    \end{equation}

    \item \textit{F1 score:} The F1 score is a statistical measure calculated as the harmonic mean of precision 
and recall values. It is a popular metric for evaluating binary classification models, especially with an imbalanced dataset.
    
    \begin{equation}
        F1 = 2\times \frac{precision \times recall}{precision + recall} \nonumber
    \end{equation}
    
\end{enumerate}


We report the performance metrics of our model on every CWE in our threat model in Table~\ref{tab:individual_performance}. We show the results of only one exploit per CWE in Table~\ref{tab:individual_performance}. We also denote the CWEs that belong to the OWASP top-10 vulnerabilities. 

In Table~\ref{tab:individual_performance}, we observe that the accuracies of all the models are quite high in comparison to other metrics. This is due to an imbalanced dataset. Thus, precision, recall, and F1 scores are more reliable metrics for evaluating our model on this dataset than accuracy. We observe that ML-FEED achieves the best F1 scores for most CWEs. The transformer performs better than ML-FEED for a few CWEs. However, we observe later in Table~\ref{tab:overheads} that the transformer incurs orders of magnitude higher computation overhead than ML-FEED. This makes ML-FEED more practical in real-world scenarios. 

\subsubsection{Error analysis} 
Here, we analyze the shortcomings of ML-FEED. ML-FEED outputs false positives and false negatives for some of the CWEs. We observe in Table~\ref{tab:individual_performance} that all the state-of-the-art ML models perform poorly on
CWE-319. CWE-319 vulnerabilities arise due to the storage of sensitive information in cleartext. One must
manually check the repository for sensitive content or train separate NLP models to catch sensitive data in
cleartext to detect this vulnerability. We aim to tackle this problem in the future by cascading a sensitivity-predicting NLP model with ML-FEED. In Table~\ref{tab:individual_performance}, we also observe that ML-based methods have a relatively low recall for detecting CWE-400 vulnerabilities. CWE-400 is a denial-of-service (DoS) attack. DoS attacks are 
difficult to prevent because they are highly dependent on the system architecture. Thus, system-specific intrusion detection systems are more efficient in detecting DoS attacks. However, the high precision scores achieved by ML-based methods in Table~\ref{tab:individual_performance} can help improve the efficiency of intrusion detection systems~\cite{saha2021sharks, saha2021machine5g}.

We can use the precision, recall, and F1 score metrics to evaluate only binary classification problems. However, our exploit classification task is a multi-label classification problem with $79$ labels. For a multi-label classification task with $n$ labels, we have $n$ confusion matrices. Precision, recall, and F1 scores cannot handle multiple confusion matrices by design. However, we can compute these metrics for each of the $n$ confusion matrices and average them to obtain the mean precision, recall, and F1 scores. 
We show the mean performance metrics of ML-FEED and state-of-the-art sequence models across all 79 labels in Table~\ref{tab:multi-label_performance}. In Table~\ref{tab:individual_performance}, we observed that accuracy is not a reliable performance metric for our dataset. Thus, we do not report the average accuracy in Table~\ref{tab:multi-label_performance}.

\begin {table}[h]
\centering
\vspace{0.5cm}
\caption {Performance of ML models on 79 exploits. All values depict percentages (\%).}
\begin{tabular}{|c|c|c|c|}
\hline
\textbf{Model} & \textbf{Precision} & \textbf{Recall} & \textbf{F1}\\ \hline
LSTM & 96.4 & 93.6 & 94.9\\ \hline
Transformer & 97.6 & 96.1 & 96.8\\ \hline
ML-FEED & 98.2 & 97.4 & 97.8\\ \hline
\end{tabular}
\vspace{-0.2cm}
\label{tab:multi-label_performance}		
\end {table}

\subsubsection{Computation overhead}
We demonstrate the computation overheads of each model in terms of their inference time and model size in 
Table~\ref{tab:overheads}. Higher overheads degrade the applicability of the models in real-world scenarios.

\begin {table}[h]
\centering
\vspace{0.3cm}
\caption {The inference time and model size of ML models}
\begin{tabular}{|c|P{1.7cm}|P{1.8cm}|P{1.5cm}|}
\hline
\textbf{Model} & \textbf{Inference time ($\mu$s)} & \textbf{Number of parameters} & \textbf{Model size}\\ \hline
LSTM & 475.81 & 213,116 & 2.5 MB\\ \hline
Transformer &  495162.72 & 117M & 523 MB\\ \hline
ML-FEED & 6.53 & 46,788 & 970.5 KB\\ \hline
\end{tabular}
\label{tab:overheads}		
\end {table}

Table~\ref{tab:overheads} shows that ML-FEED is $72.9\times$ and $75828.9\times$ faster than the lightweight LSTM 
and GPT-2 models, respectively. It also has $4.5\times$ and $2500.6\times$ fewer parameters than the LSTM and 
GPT-2 models, respectively. 

\begin{figure}[h]
    \centering
    \includegraphics[scale=0.41]{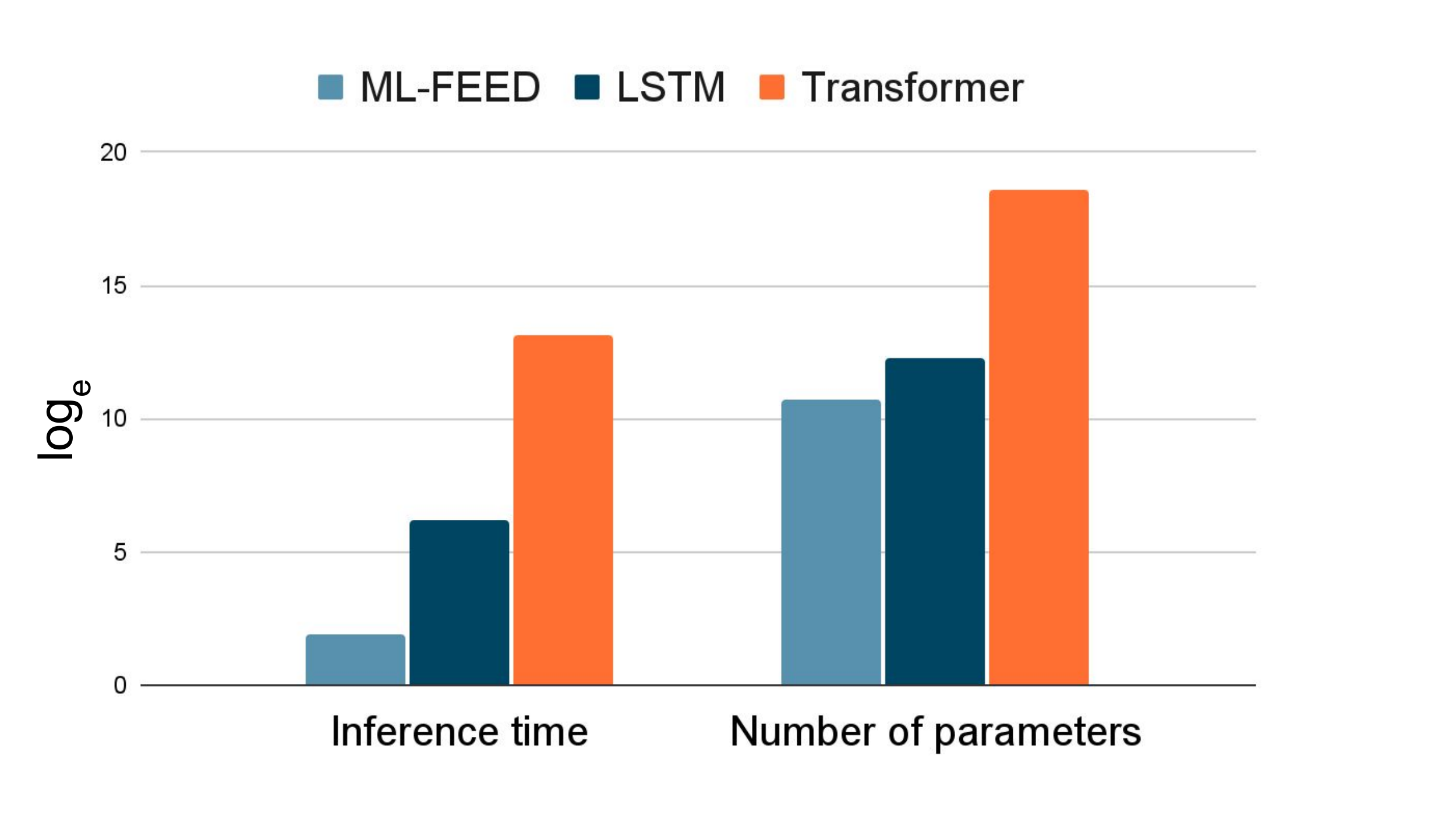}
    \caption{Comparison of the efficiency of ML-FEED with state-of-the-art ML-based sequence models}
    \label{fig:efficiency}
\end{figure}

In Fig.~\ref{fig:efficiency}, we observe that the LSTM is more efficient than the transformer. The transformer
is inefficient because it analyzes a sequence of the last 1024 API calls for inference at every time step.
However, the LSTM analyzes only two fixed-length vectors at every time step: the current API call and the
\textit{hidden state}. Fig.~\ref{fig:efficiency} also shows that ML-FEED is more efficient than the LSTM. This is because it does not maintain and update a hidden state vector at every time step. Instead, ML-FEED uses a state-table to keep track of the sequence of API calls.

\begin {table}[]
\vspace{0.3cm}
\centering
\caption {Vulnerability detector feature comparison} 
\label{tab:feature_comparison}
\scalebox{1}{
\begin{tabular}{|p{2.7cm}|p{0.9cm}|p{0.9cm}|p{0.9cm}|p{0.9cm}|}
\hline
\textbf{Model} & \textbf{Run-time} & \textbf{Rule-based} & \textbf{ML-based} & \textbf{Efficient} \\ \hline

Constraint solving \cite{thome2017search} & \cmark & \cmark & \xmark & \cmark \\ \hline
Modular checking \cite{hackett2006modular}& \cmark & \xmark & \xmark & \cmark \\ \hline
VulDeePecker \cite{li2018vuldeepecker} & \xmark & \xmark & \cmark & \xmark \\ \hline
$\mu$-VulDeePecker \cite{zou2019muvuldeepecker} & \xmark & \xmark & \cmark & \xmark \\ \hline
Yamaguchi et al. \cite{yamaguchi2011vulnerability} & \xmark & \xmark & \cmark & \cmark \\ \hline
SHARKS \cite{saha2021sharks} & \cmark & \xmark & \cmark & \xmark \\ \hline
GRAVITAS \cite{brown2021gravitas} & \cmark & \xmark & \cmark & \xmark \\ \hline
MLASA \cite{saha2021machine5g} & \cmark & \xmark & \cmark & \xmark \\ \hline
GPT-2 \cite{csahin2021malware} & \xmark & \xmark & \cmark & \xmark \\ \hline
SySevVR \cite{li2021sysevr} & \xmark & \xmark & \cmark & \xmark \\ \hline
Russell et al. \cite{russell2018automated} & \xmark & \xmark & \cmark & \cmark \\ \hline
LSTM \cite{xu2018vulnerability} & \xmark & \xmark & \cmark & \xmark \\ \hline
\textbf{ML-FEED} & \cmark & \cmark  & \cmark & \cmark \\ \hline

\end{tabular}}	
\end {table}

\subsection{Comparison}

Table~\ref{tab:feature_comparison} shows that ML-FEED is the only framework that satisfies all the features in the table. ML-FEED can perform run-time analysis effectively for all exploits in its training set. The rule-based fingerprint detector in the sequence monitor makes it more efficient than other state-of-the-art ML methods. The exploit classifier in the inference pipeline enables pattern-based exploit detection to detect a wide range of exploits.

\subsection{Evaluation with Real Attack Traces}
\label{sec:libraries}
To demonstrate the practical usage of our model, we use ML-FEED to evaluate three popular open-source libraries with existing CVEs. We choose open-source libraries because our model requires access to the source code to analyze potential vulnerabilities. 
We evaluate ML-FEED on the attack traces of \textit{CVE-2021-25646}, \textit{CVE-2021-39139}, and \textit{CVE-2020-1958}. ML-FEED successfully identifies all three vulnerabilities. We describe the detected vulnerabilities next.

\subsubsection{Remote code execution} ML-FEED identifies two remote code execution (RCE) vulnerabilities in the 
popular Apache Druid and XStream library. Apache Druid is a popular library that provides real-time 
database support for powering modern analytics applications. On the other hand, XStream allows users to serialize 
objects to XML and back again. However, while evaluating Apache Druid, ML-FEED detects an RCE 
vulnerability that allows a malicious attacker to execute JavaScript codes embedded in various types of requests. 
An authenticated user can send a specially-crafted request, regardless of server configuration, which forces Druid to run user-provided JavaScript code for that request. 
This vulnerability is reported in 
\textit{CVE-2021-25646}
, with a high severity score of 8.8. ML-FEED identifies another RCE vulnerability that may 
allow a remote attacker to load and execute arbitrary code from a remote host just by manipulating the processed 
input stream. This vulnerability has been reported in \textit{CVE-2021-39139}
, with a high severity score of 8.5.

\subsubsection{Lightweight directory access protocol (LDAP) injection} ML-FEED also identifies an LDAP injection vulnerability in the Apache Druid library. This vulnerability allows Druid API callers to bypass the \textit{credentialsValidator.userSearch} filter 
barrier that determines if a valid LDAP user is allowed to authenticate with Druid. Callers can also retrieve any LDAP attribute of users from the LDAP server, as long as that information is visible to the Druid server. This information disclosure does not require the caller to be a valid LDAP user. 
This vulnerability is reported in \textit{CVE-2020-1958}~\cite{cve20201958}
, with a medium severity score of 6.5.

\section{Related Work}
\label{sec:related_work}

Vulnerability exploit detection is a fundamental challenge in software security, for which researchers have 
proposed many notable solutions. The two broad categories of classical vulnerability detection approaches are rule-based and similarity-based methods. Rule-based methods involve manually crafting the rules of 
vulnerabilities~\cite{hackett2006modular, tv-puf, thome2017search, al2019oauthlint}. These rules attempt to encompass 
both the signatures and behaviors of the exploits~\cite{aaraj2008dynamic}. The vulnerabilities and exploit 
payloads are detected by naive pattern matching against the rules. As discussed in Section~\ref{sec:Methodology}, 
naive pattern matching suffers from generalization limitations and considerable time 
overheads~\cite{aaraj2008dynamic}. ML-FEED overcomes these limitations by using ML and reducing the search space 
for vulnerabilities. Similarity-based approaches rely on code sharing and code cloning between applications. The 
assumption is that the exposure is replicated if a vulnerable code is cloned. These approaches aim to find 
similarities with vulnerable codebases. 
Similarity-based vulnerability detection techniques are effective only when codebases are cloned into the testing project.

Vulnerability detection methods have undergone a radical transformation with the advent of 
ML~\cite{saha2021sharks, saha2021machine5g,
brown2021gravitas, saha2022system}. ML-based methods perform much better than rule-based methods. Although ML can accurately learn 
the underlying patterns of vulnerabilities, manual effort is still required to characterize the vulnerable 
programs. VulDeePecker~\cite{li2018vuldeepecker} and $\mu$VulDeePecker~\cite{zou2019muvuldeepecker} are among the 
first fully-automated ML models to detect and classify vulnerabilities successfully. Sequence models like transformers and LSTMs have found success in modeling exploit chains and detecting exploits. However, 
they suffer from degraded performance in capturing long-range dependencies in the attack sequences. Transformers are better at capturing long-range dependencies than RNNs. Hence, they have found various applications in cybersecurity analysis~\cite{wu2021literature}. However, transformers incur large computation overheads, limiting their application in real-world scenarios. 

Recent methods like ATLAS~\cite{alsaheel2021atlas} and ALchemist~\cite{yu2021alchemist} have extended the applications of sequence-based ML analysis to domains like advanced persistent threat analysis and audit report analysis-based attack investigation. Malware detection is another domain in which ML has been highly successful~\cite{lucas2021malware,zhang2020enhancing, chen2020training}. ML has also shown promising results in detecting attacks on networks~\cite{ye2021netplier}, operating systems~\cite{zhang2021osprey}, and web applications~\cite{bai2021runtime}. Despite the rapid progress of ML methods, they suffer from a lack of explainability. ML algorithms generally work as black boxes.  Hence, security analysts cannot easily interpret their predictions. This lack of transparency diminishes the trust analysts place in ML models. Fortunately, researchers have recently developed explainable ML models for cybersecurity~\cite{ganz2021explaining, pirch2021tagvet, warnecke2020evaluating, wang2021exposing} that can potentially increase the confidence that can be placed in ML models.

\section{Conclusion}
\label{sec:conclusion}
In this article, we presented ML-FEED, a lightweight ML-based exploit detection framework that is orders of 
magnitude faster than state-of-the-art methods while achieving better performance. It analyzes sequences of API 
calls and raises an alarm when an exploit is executed. It can also classify the exploit into CWE categories. We 
also proposed an automated method to extract intelligence from the CWE and CVE vulnerability databases, thus 
mitigating the need for human intervention. Finally, we used ML-FEED to analyze three public libraries and found the existing CVEs in all of them.
Finally, ML-FEED was successful in detecting existing CVEs in three public libraries.

\section*{Acknowledgment}
We thank Bhishma Dedhia for his help with implementing the GPT-2 transformer model.

\bibliographystyle{IEEEtran}
\bibliography{output}

\end{document}